\documentclass[12pt,epsfig]{article}
\usepackage{amsmath}
\usepackage{graphicx}
\usepackage{epstopdf}
\usepackage{lscape}
\usepackage{longtable}
\usepackage[usenames, dvipsnames]{color}
\usepackage{bookmark}

\begin{document}
	\begin{center}
		\textbf {A new measure for assessment of clustering based on kernel	density estimation}\\
	\end{center}
	\begin{center}
		Soumita Modak$^{*}$\\
		Department of Statistics\\ Basanti Devi College\\ affiliated with \\University of Calcutta\\
		147B, Rash Behari Ave, Kolkata- 700029, India\\
		email: soumitamodak2013@gmail.com\\
		Orcid id: 0000-0002-4919-143X
		
	\end{center}
	\begin{abstract}
		A new clustering accuracy measure is proposed to determine the unknown number of clusters and to assess the quality of clustering of a data set given in any dimensional space. Our validity index applies the classical nonparametric univariate kernel density estimation method to the interpoint distances computed between the members of data. Being based on interpoint distances, it is free of the curse of dimensionality and therefore efficiently computable for high-dimensional situations where the number of study variables can be larger than the sample size. The proposed measure is compatible with any clustering algorithm and with every kind of data set where the interpoint distance measure can be defined to have a density function. Our conducted simulation study proves its superiority over widely used cluster validity indices like the average silhouette width and the Dunn index, whereas its applicability is shown with respect to a high-dimensional Biostatistical study of Alon data set and a large Astrostatistical application of time series with light curves of new variable stars.
	\end{abstract}
	
	
	\section{Introduction}
	Cluster analysis is the unsupervised classification procedure to group a data set into homogeneous clusters such that similar members of the data constitute a meaningful class representing a source of generation different compared to the other classes. Various clustering algorithms are available in the literature (Bezdek, 1981; Jain et al., 1999; McLachlan and Peel, 2000; Kaufman and Rousseeuw, 2005; Arias-Castro et al., 2016; Matioli et al., 2018; Modak et al., 2018, 2020, 2022; Modak, 2019; Modak, 2021a). Choice of the clustering method is problem specific, while model-based methods depend on the underlying assumptions about the distribution of the sample, nonparametric methods are also sensitive to the shapes of the resulting clusters (Kaufman and Rousseeuw, 2005; Modak, 2021b). Therefore, it is always important to assess the quality of the clustering results by an efficient accuracy measure which helps to determine the unknown value of the number of groups ($K$) present in the data set. In fact, the same clustering method could indicate different values of $K$ for various cluster validity indices.
	
	In this regard, we consider a novel measure for assessment of clustering based on the classical nonparametric univariate kernel density estimation method (Silverman, 1986), which can evaluate the quality of cluster analysis of a data set with inherent grouping (i.e. $K>1$ is assumed) obtained through any hard clustering algorithm wherein every data member is classified in only one group. It is useful for data measured on arbitrary scales for which the appropriately adopted interpoint distance measure (may not be strictly metric; Kaufman and Rousseeuw, 2005; Modak et al., 2020) can be defined to have a probability density function (unknown). Our validity index applies the classical nonparametric univariate kernel density estimation method to the interpoint distances computed between the members of data so that our procedure does not need to know the distribution of the original sample or the density function of their interpoint distances. Being based on interpoint distances, it is free from the curse of dimensionality and therefore efficiently computable for high-dimensional situations where the number of study variables can be larger than the sample size. Our proposed measure lying between -1 to 1 indicates better clustering with a larger value. It can be used to estimate the unknown true value of $K$ for a given data set by assessing the quality of cluster analysis obtained through an unsupervised classification algorithm, or to compare the clusterings of data members from various classification methods for given a value of $K$. Our simulation study on multivariate, mixed and complex (where multivariate dependence structure is specified by a copula; see, for example, Nelsen, 2006; Modak and Bandyopadhyay, 2019) high-dimensional data sets 
	shows that the novel measure surpasses the widely used cluster validity indices, namely the average silhouette width (Rousseeuw, 1987; Kaufman and Rousseeuw, 2005) and the Dunn index (Dunn, 1974). Two interesting data sets, one involving high-dimensional Biostatistical application of Alon data (Alon et al., 1999) and another one concerning big time series data from Astrostatistical study on challenging light curves of new variable stars found in the Milky way (Modak et al., 2020, 2022), are used to prove its usefulness.

	The paper is organized in the following fashion. Section 2 depicts the proposed clustering assessment measure in detail along with its rival measures under consideration. Section 3 explains the simulation study, whereas Section 4 shows its application to real data sets with Section 5 concluding.
	
	\section{Method}\label{method}
	For a real-valued random sample $X_1,\ldots,X_n$ of size $n$ from
	a continuous univariate distribution with unknown probability density function
	$f$, we consider the nonparametric kernel density estimator $\hat{f}$ given by
	\begin{equation}\label{kde}
		\hat{f} (x) = \frac{1}{nh} \sum\limits_{i=1}^{n}Ker\bigg(\frac{x-X_i}{h}\bigg),
	\end{equation}
	where $h$ is the smoothing parameter and $Ker(\cdot)$ is a real-valued kernel function with $\int_{-\infty}^{\infty} Ker(x) dx=1$ (Silverman, 1986; Bandyopadhyay and Modak, 2018). In this paper, we implement the Gaussian kernel with 
	\begin{equation}\label{GK}
		Ker(x)=\frac{1}{\sqrt{2\pi}}\exp\bigg(-\frac{x^2}{2}\bigg).
	\end{equation}
	Then, under the assumption that $f$ makes a unimodal distribution, we can obtain an estimate of the mode using Eq.~\eqref{kde} as  
	\begin{equation}
		\hat{m}=\bigg\{\hat{m}\in \Re: \underset{x\in \Re}{\max}\hspace{.05 in}\{\hat{f}(x)\}=\hat{f}(\hat{m})\bigg\},
	\end{equation}
	where $\Re$ stands for the real line; however, in practical situations, $\hat{m}$ is found within a reasonable interval $\subset \Re$ dependent upon the given sample. 
	
	Now, to design our validity index, we apply the above-mentioned mode estimation procedure to the interpoint distances, assumed to possess a probability density function (unknown), between the data members for assessment of clustering them in $K(>1)$ mutually exclusive and exhausted clusters $C_1$,\ldots,$C_K$ of sizes $n_1,\ldots,n_K$, respectively, with $\sum_{k=1}^{K}n_k=n$ as the total number of members in the given data set. Use of the interpoint distances enables us to extend the application of the univariate kernel density estimation technique to multivariate data set in arbitrary dimensional space.
	
	For the $m-$th member of the $k-$th cluster $C_k$ or the corresponding observation (univariate or multivariate) $M_{k,m}$, $m=1,\ldots,n_k$, $k=1,\ldots,K$, with $d(M,M')$ representing the interpoint distance (may not be strictly metric) between any members $M$ and $M'$, we proceed as follows. \\
	(1) Estimate the mode as  $\hat{m}_{k,m}$ for the sample of interpoint distances between the member $M_{k,m}$ and all other members of its cluster $C_k$: $\{d(M_{k,m},M_{k,m'}), m'$ $(\neq m)=1,\ldots,n_{k}\}$.\\
	(2) Estimate the mode as $\hat{m}_{k';k,m}$ for the sample of interpoint distances between the member $M_{k,m}$ and all members of other cluster $C_{k'}$ with $k' \neq k$: $\{d(M_{k,m},M_{k',m'}), m'=1,\ldots,n_{k'}\}$.\\
	(3) Repeat step (2) for all other clusters and compute $\hat{m}_{k';k,m}$ for $k'(\neq k)=1,\ldots,K$.\\
	(4) We define the nearest cluster $C_{nc}$ to the member $M_{k,m}$ with the smallest of sample modes from step (3), i.e.
	\begin{equation}\label{nc}
		\underset{1\leq k'(\neq k)\leq K}{\min}\hspace{.05 in}\bigg\{\hat{m}_{k';k,m}\bigg\}=\hat{m}_{nc;k,m}\hspace{.1 in}.
	\end{equation} 
	Computationally considering, if Eq.~\eqref{nc} holds for more than one $C_{k'}$ in the given sample, then $C_{nc}$ is selected randomly from them.\\ 
	(5) Calculate 
	\begin{equation}
		m(M_{k,m})=\frac{\hat{m}_{nc;k,m}-\hat{m}_{k,m}}{\max\bigg\{\hat{m}_{nc;k,m},\hat{m}_{k,m}\bigg\}}\hspace{.05 in}\text{for}\hspace{.05 in}m=1,\ldots,n_k; k=1,\ldots,K.
	\end{equation}
	If $M_{k,m}$ is the only member of the cluster $C_k$, then $\hat{m}_{k,m}$ is not available and we define $m(M_{k,m})=0$. On the other hand, if $C_k$ has only two members, i.e. one existing member belonging to $C_k$ other than $M_{k,m}$, then $\hat{m}_{k,m}=d(M_{k,m},M_{k,m'})$ with $m'\neq m$. Similarly, for any other cluster $C_{k'},k'\neq k$ having single member $M_{k',m'}$, we reasonably choose $\hat{m}_{k';k,m}=d(M_{k,m},M_{k',m'})$.\\ 
	(6) Our clustering accuracy measure is defined as
	\begin{equation}\label{measure}
		M_{clus}=\frac{1}{n}\sum\limits_{k=1}^{K}\sum\limits_{m=1}^{n_k}m(M_{k,m}),
	\end{equation}
	which takes values in $(-1,1)$ with better clustering indicated by a larger value.
	\subsection{Motivation}
	For each given member $M_{k,m}$, the set of interpoint distances:
	$\{d(M_{k,m},M_{k,m'}),$ $m'(\neq m)=1,\ldots,n_{k}\}$ constitutes a random sample of size $n_{k}-1$ from a univariate continuous distribution whose mode is estimated as $\hat{m}_{k,m}$ using the univariate kernel density estimation method. 
	Likewise, the set of interpoint distances:  $\{d(M_{k,m},M_{k',m'}), m'=1,\ldots,n_{k'}; k'\neq k\}$ forms a random sample of size $n_{k'}$ coming from a univariate continuous distribution with estimated mode $\hat{m}_{k';k,m}$ obtained through the kernel density estimation procedure.
	
	Now, we can theoretically visualize that for given $M_{k,m}$ conditionally, $S_1=\{d(M_{k,m},M_{k,m'}), m'(\neq m)=1,\ldots,n_{k}\}$ is a random sample from a univariate population with continuous distribution function $F(t|M_{k,m})=F(t)$ (say) and $S_2=\{d(M_{k,m},M_{nc,m'}), m'=1,\ldots,n_{nc}; nc\neq k\}$ is another random sample coming independently of $S_1$ from a univariate population with continuous distribution function $G(t|M_{k,m})=G(t)$ (say). Then, in the context of nonparametric statistical analysis, the assessment of clustering the member $M_{k,m}$ in the cluster $C_k$ can be explained with the help of the location-shift model:
	\begin{equation*}
		G(t)=F(t-\theta)\hspace{.05in} \text{for\hspace{.05in}all}\hspace{.05in} t,
	\end{equation*}
	where $\theta$ is the difference in locations of the two populations.
	
	If the member $M_{k,m}$ clustered in $C_k$ originally belongs to the population of the cluster $C_k$, then the interpoint distances from the set $S_2$ are likely to be stochastically larger than the distances of the set $S_1$, under which circumstances we expect $\theta$ to be significantly greater than 0.
	If membership of the member $M_{k,m}$ in its assigned cluster $C_k$ is equally likely to its membership in the nearest cluster $C_{nc}$ as defined in Eq.~\eqref{nc}, then $\theta$ is expected to be close enough to 0 as the interpoint distances of the sets $S_1$ and $S_2$ tend to be close; whereas $\theta$ being significantly smaller than 0 is indicative of $M_{k,m}$ coming from the distribution of $C_{nc}$, in which situation interpoint distances of $S_1$ are expected to be stochastically larger than those from $S_2$.   
	
	The location shift parameter $\theta$ could be studied as the difference between the population modes, which are estimated here as $\hat{m}_{k,m}$ and $\hat{m}_{nc;k,m}$ using kernel density estimators. Therefore, the above three situations now can respectively expect to possess $\hat{m}_{nc;k,m}$ quite larger than, close enough to or much smaller than $\hat{m}_{k,m}$.
	Consequently, $m(M_{k,m})$ is greater than, almost equal to, or less than 0. By construction, each $m(M_{k,m})$ lies between -1 to 1, wherein evidently larger is the value better is the clustering. A value of $m(M_{k,m})$ close to 1 implies that $\hat{m}_{k,m}$ is much smaller than $\hat{m}_{nc;k,m}$ describing the fact that member $M_{k,m}$ is well clustered in $C_k$; whereas $m(M_{k,m})$ having a value near -1 indicates the opposite, where member $M_{k,m}$ is closer to its nearest cluster $C_{nc}$ than to the cluster $C_k$ it is allocated to in the cluster analysis and therefore it is concluded that $M_{k,m}$ is badly clustered in $C_k$. On the other hand, an approximate 0 value for $m(M_{k,m})$ means the member $M_{k,m}$ can belong either to the assigned cluster $C_k$ or to its nearest cluster $C_{nc}$, because $M_{k,m}$ lies somewhere in between the two clusters. Finally, designed as the arithmetic mean of $\{m(M_{k,m}),\hspace{.05in}m=1,\ldots,n_k; k=1,\ldots,K\}$, the same interpretation is true for our proposed measure $M_{clus}$, which now assesses the overall quality of clustering for all the members given in the data set.
	
	It is to be noted that our clustering assessment measure $M_{clus}$ is a function of $K$, however, for simplification, we avoid incorporating such $K$ dependent notation. For a given value of $K$, $M_{clus}$ lying between -1 to 1 indicates better classification with a higher value while comparing different clustering algorithms applied to the same data; whereas for a data set with unknown value of $K$, we compute $M_{clus}$ at different values for $K$ to estimate the true value of $K$ as $\hat{K}$ such that
	\begin{equation}
		\hat{K}=\bigg\{\hat{K}\in D: \underset{K\in D}{\max}\hspace{.05 in}\{M_{clus}(K)\}=M_{clus}(\hat{K})\bigg\} \hspace{.05 in}, 
	\end{equation}
	where numerically $D=\{2,3,...,n-1\}$, however, practically $\max(D)$ is considered, depending on the data under study, to be much smaller than the size of the data set $n$. 
	\subsection{Competitors} 
	We use the following two well-known clustering accuracy measures for comparison.
	
	Average silhouette width: For each member $M_{k,m}$, let
	$a(M_{k,m})$ be the average of interpoint distances of $M_{k,m}$ to all other members
	of its own cluster and $b(M_{k,m})$ be the minimum of average interpoint distances of $M_{k,m}$ to members of other clusters. Then, the silhouette width for the member $M_{k,m}$, $s(M_{k,m})$ is defined as
	\begin{equation}
		s(M_{k,m})=\frac{b(M_{k,m})-a(M_{k,m})}{\max\bigg\{a(M_{k,m}),b(M_{k,m})\bigg\}}.
	\end{equation}
	The average silhouette width is (Rousseeuw, 1987; Kaufman and Rousseeuw, 2005)
	\begin{equation}
		\text{ASW}=\frac{1}{n}\sum\limits_{k=1}^{K}\sum\limits_{m=1}^{n_k}s(M_{k,m}),
	\end{equation}
	ASW $\in [-1,1]$ whose larger value suggests better classification.
	
	Dunn Index: It is designed to be the ratio of the smallest of interpoint distances between the members belonging to different clusters to the largest of interpoint distances between the members belonging to the same cluster as follows (Dunn, 1974):
	\begin{equation}
		\text{Dunn}=\frac{\underset{k\neq k'}{\underset{1\leq k,k' \leq K}{\min}}\bigg\{ \underset{ 1\leq m \leq n_k, 1\leq m' \leq n_{k'}}\min d(M_{k,m},M_{k',m'})\bigg\}}{\underset{1\leq k \leq K}\max \bigg\{\underset{1\leq m,m' \leq n_k}\max d(M_{k,m},M_{k,m'})\bigg\}}.
	\end{equation}
	Dunn $\in (0,\infty)$ indicates better clustering with a higher value.\\
	\section{Simulation study}  
	Efficacy of our measure is demonstrated in terms of simulation study, where we draw samples independently from $K'$ different populations with each population representing one cluster. We then combine the samples into one working data set and apply a clustering algorithm, where for different values of $K\in D=\{2,3,\ldots,6\}$ we compute the values of our clustering accuracy measure along with its competitors. The optimal values of these validity indices independently estimate the value of $K$ for the given sample as $\hat{K}$ which is now to be compared with the true known value of $K$, i.e. the number of populations under study which is $K'$ here. Under $K'$ different populations specified, the samples are drawn `$Rep$' times and the above-mentioned procedure of determining $\hat{K}$ is repeated for each realization $r=1,\ldots,Rep$, where performance of a measure is evaluated by 
	\begin{equation}\label{comparison}
		p=\bigg\{\sum\limits_{r=1}^{Rep}I^{(r)}_{K'}(\hat{K})\bigg{/}Rep\bigg\}\times100 
	\end{equation} 
	with
	\[I^{(r)}_{K'}(\hat{K})= \begin{cases} 
		\hspace{.13in}1 & \text{if}\hspace{.13in} \hat{K}=K',\\
		\hspace{.13in}0 &  \text{if}\hspace{.13in} \hat{K}\neq K',
	\end{cases}
	\]
	in the $r-$th realization.
	We consider $Rep=100$. Comparing relative performance, a measure is better with a higher value for $p$, which gives the percentage of efficacy of the corresponding measure in identifying the true number of clusters present in the given data set.   
	
	Using the nonparametric density estimator from Eq.~\eqref{kde} with the Gaussian kernel in Eq.\eqref{GK}, we consider 
	$h=$ $1.06$ $\sigma$ $n^{- 1/ 5}$ for which the asymptotic mean integrated squared error is minimized while estimating a normal density function with standard deviation $\sigma$ (Silverman, 1986; Matioli et al., 2018). Here $\sigma$ is estimated as $\hat{\sigma}$ by the standard deviation of the sample of interpoint distances and the corresponding value of $h$ is denoted by $h^*=$ $1.06$ $\hat{\sigma}$ $n^{- 1/ 5}$. 
	Followings are the different set-ups of simulation under study, where for comparison throughout a particular set-up we use the same distance measure for implementing the clustering algorithms and computing the different cluster validity indices. We denote the distinct simulation scenarios by S1, S2 and S3.
	
	S1) Firstly, we consider a multivariate normal population with mean vector $\boldsymbol{\mu}$ and dispersion matrix $\Sigma$ in $v-$dimensional space denoted by $N_v(\boldsymbol{\mu},\Sigma)$. We draw random samples each of size 50 independently 
	from three (i.e. true value of $K=K'=3$) different populations $N_{10}(\boldsymbol{\mu_i},\Sigma)$, $i=1,2,3$, where $\boldsymbol{\mu}_i=(\mu_i,\ldots,\mu_i)'$ with $\mu_1=-3,\mu_2=0,\mu_3=3$ and $\Sigma=(\sigma_{ij})_{i,j=1,...,10}$ with $\sigma_{ii}=1$ $\forall i$ and $\sigma_{ij}=0.5$  $\forall i\neq j$. We combine the samples to form our working data set, where
	interpoint distances are calculated using the Euclidean metric, and apply the agglomerative hierarchical algorithm with average linkage (Kaufman and Rousseeuw, 2005). The results in terms of $p$ (see, Eq.~\ref{comparison}) are reported in Table~\ref{t:t1}.
	
	S2) Secondly, we apply $K-$medoids clustering algorithm (Kaufman and Rousseeuw, 2005; Modak et al., 2017; Modak et al., 2020) with the Gower's distance (Gower, 1971; we implement an extended version proposed in Kaufman and Rousseeuw, 2005) to identify the two inherent clusters consisting of mixed type of bivariate data having binary and continuous variables. For one population, the first variable is taking the values 0 or 1 with respective probabilities 0.8 and 0.2, and the second variable is following, independently of the former, a Cauchy distribution with location 0 and scale 1; whereas for the other population, values 0 and 1 are assigned to the first variable with probabilities 0.2 and 0.8 respectively, and the second variable is independently having a Cauchy distribution with parameters specified by location at 3 and scale at 1. Results based on random samples of sizes 45 from each of the two populations are given in Table~\ref{t:t1}.
	
	S3) Lastly, we simulate data from a high-dimensional space with the number of variables under study much higher than the sample size, where a sample having 500-dimensional data vectors of size 150 is drawn from the multivariate population with marginal distributions $LN(0,0.8)$, i.e. each variable follows a log-normal distribution whose natural logarithm has mean 0 and standard deviation 0.8. Here dependence among the variables is established by a normal copula (Nelsen, 2006) characterized by $N_{500}(\boldsymbol{\mu},\Sigma)$ where $\boldsymbol{\mu}=(0,\ldots,0)'$ and $\Sigma=(\sigma_{ij})_{i,j=1,...,500}$ with $\sigma_{ii}=1$ $\forall i$ and $\sigma_{ij}=0.75$  $\forall i\neq j$. Let $F$ denote the distribution function of $N_{500}(\boldsymbol{\mu},\Sigma)$ where $F_i$ be the marginal distribution function for the $i-$th variable with inverse function $F_i^{-1}$, then the normal copula is expressed as
	\begin{equation*}
		C(u_1,\ldots, u_{500}) = F\{F_1^{-1} (u_1),\ldots,F_{500}^{-1}(u_{500})\}, 0<u_1,\ldots,u_{500}<1.
	\end{equation*}
	The difference of location is created by every variable-wise addition of 3 and -3 to the first 50 and last 50 data vectors, respectively. Thus, three clusters are formed and we perform $K-$means clustering method on this data set as a whole using the Hartigan--Wong algorithm (Hartigan and Wong, 1979) with the Euclidean metric; subsequently we compute the cluster validity indices whose results over 100 realizations are given in Table~\ref{t:t1}. 
	
	In all the considered situations, our cluster validity index outperforms the other two (see, Table \ref{t:t1}). 
	\section{Real-life applications}
	Our first real data set involves a high-dimensional application from Biostatistics, known as `Alon data' (Alon et al., 1999), which consists of 2000 genes as study variables measured on 62 patients as members of the sample with two inherent clusters having 40 patients diagnosed with colon cancer and 22 healthy patients. We perform cluster analysis on this data set using the agglomerative hierarchical algorithm with Ward's criteria (Murtagh and Legendre, 2014) and Euclidean norm. Our measure $M_{clus}$ (with $h=h^{*}$) correctly estimates $\hat{K}=2$, ASW also indicates the same but with quite less precision because the values of ASW computed for $K=2$ and for $K=3$ are close enough, whereas Dunn gives $\hat{K}=4$ (see, Table~\ref{t:t2}). To see the robustness of our measure with respect to the choice of the bandwidth $h$, we consider
	$h=$ $1.06$ $\hat{\sigma}$ $n^{- 1/ \alpha}$ where we vary $\alpha \in \{0.1,0.1+\epsilon,0.1+2\hspace{.05in}\epsilon,\ldots,0.1+99\hspace{.05in}\epsilon=10\}$ with $\epsilon=0.1$ (Matioli et al., 2018) and find it consistently producing $\hat{K}=2$ for $M_{clus}$ values distributed with mean 0.34353 and standard error 0.00072. While the clustering method implemented here achieves 43.548\% correct classification for $M_{clus}=0.34569$ (with $h=h^{*}$), a different clustering procedure in terms of the agglomerative hierarchical algorithm with average linkage attains 56.452\% successful classification for $M_{clus}=0.41046$ (with $h=h^{*}$). It clearly exhibits the efficacy of our measure to assess the quality of a cluster analysis for the given data set, where a larger value for $M_{clus}$ represents a better clustering.  
	
	Second application concerns a large time series data from Astrostatistical study on 1318 new variable stars in the Milky Way (Miller et al., 2010; Modak et al., 2022). The observed data set on the light curves of the stars gives observations of brightness over time for each star in terms of the relative flux variation in R-band given on a continuous time scale in Heliocentric
	Julian Date. This data set is quite challenging in that the light curves are unevenly spaced of unequal lengths with observations at different times. Therefore, the requisite preprocessing of data is carried out prior to the cluster analysis (see, Modak et al., 2020). Then $K-$medoids clustering method (Kaufman and Rousseeuw, 2005; Modak et al., 2017) in association with the complexity invariance distance (Prati and Batista, 2012; Batista et al., 2014) is applied to the light
	curves each of length 272 having observations over phase interval [0,1]. It exposes two clusters of eclipsing binary stars in terms of ASW (Modak et al., 2020). Our validity index $M_{clus}$ is successful in determining $\hat{K}=2$ consistently for all values of $h=$ $1.06$ $\hat{\sigma}$ $n^{- 1/ \alpha}$ with  $\alpha \in \{1,1+\epsilon,1+2\hspace{.05in}\epsilon,\ldots,1+18\hspace{.05in}\epsilon=10\}$, $\epsilon=0.5$; whereas Dunn fails to do so producing $\hat{K}=3$.
	
	\section{Conclusion}
	A novel nonparametric accuracy measure is proposed based on the interpoint distances for the assessment of clustering. It utilizes the classical univariate kernel density estimation procedure to estimate the unknown number of clusters existent in a data set or to evaluate the quality of clustering output obtained from any classification algorithm. Its diverse use, high-dimensional applicability and great performance, demonstrated through simulation and real challenging data sets, establish it as a new strong cluster validity index.
	\section*{Acknowledgments}
		The author would like to thank the editors and the anonymous reviewer
		for their careful reading of the manuscript and appreciating the work. The author is grateful to the reviewer for intriguing inquires which helped to improve the transparency of the manuscript.
	\clearpage
	\begin{table}
		\caption{Comparison of different clustering accuracy measures in terms of $p$ for the simulated data sets}
		\begin{center}
			\begin{tabular}{cccc}
				\hline\\			
				Simulation&$M_{clus}$&ASW&Dunn\\[1ex]
				\hline\\		     			
				S1&85& 45& 11\\[1ex]
				S2&91&  74  & 45\\[1ex]
				S3&92& 50& 48\\[1ex]

				\hline
			\end{tabular}
		\end{center}
		\label{t:t1}
	\end{table}
	\clearpage
	\clearpage
	\begin{table}
		\caption{Computed values of various clustering accuracy measures for different number of clusters ($K$) as  obtained by hierarchical clustering of the Alon data set}
		\begin{center}
			\begin{tabular}{cccc}
				\hline\\			
				$K$&$M_{clus}$&ASW&Dunn\\[1ex]
				\hline\\		     			
				2&0.34569& 0.30864&0.39469\\[1ex]
				3&0.32331& 0.30826&0.39469\\[1ex]
				4&0.27929& 0.28325&0.40312\\[1ex]
				5&0.14687&0.13460&0.27928\\[1ex]	
				6&0.15348& 0.14359&0.29780\\[1ex]

				\hline
			\end{tabular}
		\end{center}
		\label{t:t2}
	\end{table}
	{}
	
\end{document}